\documentclass[twocolumn]{webofc}

\usepackage[varg]{txfonts}   
\usepackage{multirow}
\newcommand{\etal}{\textit{et al., }}
\newcommand{\eg}{\textit{e.g., }}
\newcommand{\ie}{\textit{i.e., }}
\def\vector#1{\mbox{\boldmath $#1$}}
\def\matrix#1{\mbox{\bf #1}}
\def\var{\mbox{var}}
\def\cov{\mbox{cov}}

\begin{document}
\title{Uncertainties of calculated coincidence-summing correction factors in gamma-ray spectrometry}
%
%

\author{
        \firstname{V.} \lastname{Semkova}\inst{1}\fnsep\thanks{\email{vsemkova@inrne.bas.bg}}
   \and \firstname{N.} \lastname{Otuka}\inst{2}
   \and \firstname{A.J.M.} \lastname{Plompen}\inst{3}
}

\institute{Institute for Nuclear Research and Nuclear Energy, BG-1784, Sofia, Bulgaria
\and
           Nuclear Data Section (NDS), International Atomic Energy Agency, A-1400 Wien, Austria
\and
           European Commission Joint Research Centre (EC-JRC), B-2440 Geel, Belgium
          }

\abstract{%
Uncertainty propagation to the $\gamma$-$\gamma$ coincidence-summing correction factor from the covariances of the nuclear data and detection efficiencies have been formulated.
The method was applied in the uncertainty analysis of the coincidence-summing correction factors in the $\gamma$-ray spectrometry of the $^{134}$Cs point source using a p-type coaxial HPGe detector. 

}
\maketitle
\section{Introduction}
\label{sec:introduction}
Photon spectrometry is widely applied in nuclear sciences and applications.
%
Large volume detectors and close source-to-detector geometry provide high geometrical efficiency for gamma radiation detection.
Such conditions improve the spectral output and reduce the counting time.
However, the probability for energy deposition of two or more $\gamma$-rays emitted in a cascade de-excitation increases.
It results in occurrence of true coincident summing events leading to peak intensity raise (the full energy deposited by adjacent photons is equal to the energy of another $\gamma$-line), peak intensity loss (the full energy deposition of a photon is registered with a partial deposition of other $\gamma$-rays), or new $\gamma$-lines in the spectrum.
The magnitude of the resultant true coincident summing (TCS) correction factor~\cite{KD1979,KD1988} varies greatly with the decay properties of the considered nuclide, full energy, and total efficiencies of the detector.

Various approaches to account for all possible cascade de-excitations and emitted radiation in the quantitative estimation of the TCS correction factors have been published.
Recursive formulae were proposed by Andreev et al.~\cite{DSA1973,DSA1972} for the general case of $n$ levels,
which allow simultaneous and consistent approach to all $\gamma$ transitions of a decay scheme.
In the work of McCallum and Coote~\cite{GJM1975},
the method was extended for $\beta^+$ decays.
Matrix formalism was applied to Andreev’s calculation algorithm by Semkow et al.~\cite{TMS1990}.
A technique to incorporate X-ray emission following electron capture or internal conversion was proposed in the work of M. Korun and R. Martin\v{c}i\v{c}~\cite{MK1993}.
For complex decay schemes,
Monte Carlo simulations~\cite{JPL2000} or graph theory~\cite{OS2008} have also been applied to determine the combination of the simultaneously emitted photons.

An essential part of the TCS correction factor estimation is the determination of the uncertainty in the correction factor.
Although the model equation is defined, the calculations become quite involved with increasing the number of levels and strongly depend on particular decay scheme.
In addition,
all methods applied in the calibration of the detector’s peak and total efficiencies introduce correlations between the efficiency values at different energies.
Those correlations as well as the correlations in the decay data (\eg transition probabilities) affect the uncertainty of the calculated TCS correction factor.
De~Felice et al.~\cite{PDF2000} concluded that the propagation of the uncertainty of the total efficiency results in “compression” of the uncertainty of the TCS correction for the $^{134}$Cs 604.7~keV $\gamma$-line.
Generally, the standard uncertainty propagation requires evaluation of all cascades and joint emission probabilities for the coincidence radiation based on particular decay characteristics.
The numerical calculations of the approximate values of the derivatives by
$f'(x_0)\sim[f(x_0+\Delta x)-f(x-\Delta x)]/2\Delta x$
are not very practical due to the number of variables of the model involved.
In the work of Kastlander et al.~\cite{JK2017},
the TCS correction factors and the associated uncertainties are estimated by Monte Carlo simulations of the measurement geometry and the $^{134}$Cs nuclear decay data uncertainty sampling.
The study concludes that the decay data uncertainties with magnitude of percent level could have a non-negligible impact on the TCS correction factors. 

In this work explicit formulae for the propagation of the uncertainties in the peak and total efficiencies, $\beta^-$ transition probability, $\gamma$ transition+internal conversion probability and internal conversion coefficient to the uncertainty in the TCS correction factor based on the model proposed by Andreev et al.~\cite{DSA1973,DSA1972} are presented.
The method was applied to detection of $\gamma$-ray emission following $^{134}$Cs $\beta^-$ decay.
The partial uncertainties of the TCS correction factors are presented.
The correlations of the different attributes in the ratio were taken into account as well. 

\section{Formulation}
\subsection{Correction factor~\cite{TMS1990}}
If there is no coincidence summing in $\gamma$ cascade in $n$ levels,
the probability to detect transition from level $j$ to $i$ in a precursor decay is
\begin{equation}
\left[
\sum_{p=1}^n f_p \left(
 \delta_{pj}
+x_{pj}
+\sum_{k=1}^n x_{pk}x_{kj}
+\sum_{k,l=1}^n x_{pk}x_{kl}x_{lj}
+\cdots
\right)
\right]
a_{ji}
\label{eqn:ctswocoi}
\end{equation}
with
\begin{equation*}
a_{ji}=\frac{x_{ji}}{1+\alpha_{ji}}\epsilon^p_{ji},
\end{equation*}
where $f_p$ is the $\beta$ transition probability from the precursor to level $p$,
$x_{ji}$ is the $\gamma$ transition+internal conversion probability,
$\alpha_{ji}$ is the internal conversion coefficient,
and $\epsilon^p_{ji}$ is the peak efficiency.
Kronecker's delta $\delta_{pj}$ expresses the $\beta$ transition from the precursor to level $j$.
The probabilities $f_p$ and $x_{ji}$ are normalized as
\begin{equation}
\sum_{p=0}^n f_p=1,
\quad
\sum_{i=0}^n x_{ji}=1\quad(x_{ji}=0 \mbox{ if } j\le i).
\label{eqn:norm}
\end{equation}
By denoting the decay $\gamma$ emission probability by $I_{\gamma ji}$, Eq.~(\ref{eqn:ctswocoi}) can be simplified to $I_{\gamma ji}\epsilon^p_{ji}$.
\begin{figure}[h]
\centering
\includegraphics[width=0.5 \columnwidth]{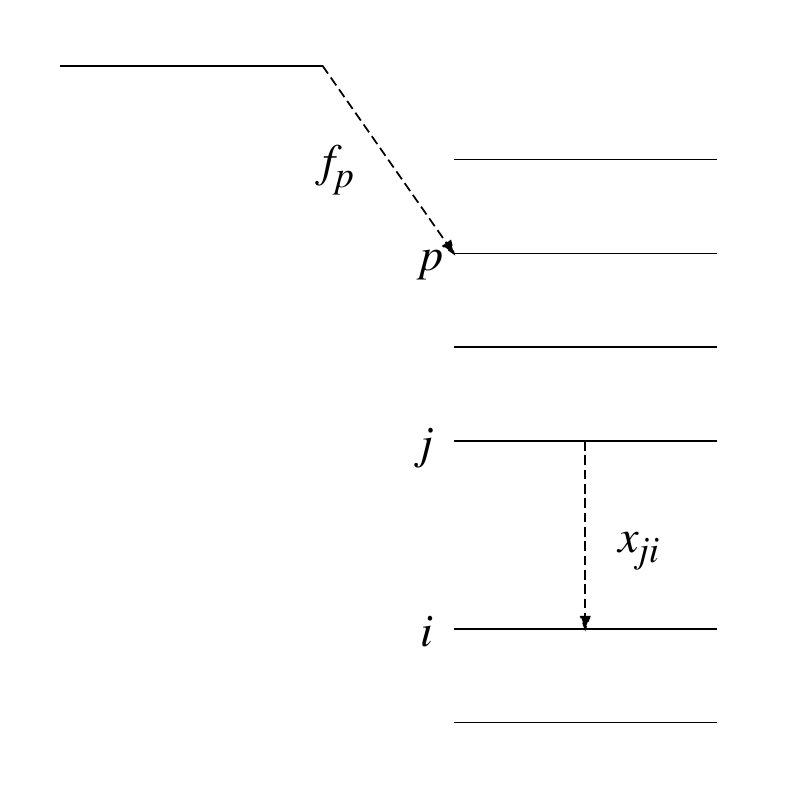}
\caption{$\beta$ transition probability $f_p$ and $\gamma$ transition plus internal conversion probability $x_{ji}$.}
\label{fig:level}
\end{figure}

The following three modifications are necessary to Eq.~(\ref{eqn:ctswocoi}) if coincidence summing is taken into account:
\begin{itemize}
\item
The first modification is for summing-out caused by detection of a transition to the level $j$ or above.
The $[\cdots]$ part of Eq.~(\ref{eqn:ctswocoi}) is modified to the probability of transition to level $j$ without any detection:
\begin{equation}
\sum_{p=1}^n f_p 
\left(
 \delta_{pj}
+b_{pj}
+\sum_{k=1}^n b_{pk}b_{kj}
+\sum_{k,l=1}^n b_{pk}b_{kl}b_{lj}
+\cdots
\right)
\label{eqn:sumout1}
\end{equation}
where 
\begin{equation*}
b_{kl}=x_{kl}-\frac{x_{kl}}{1+\alpha_{kl}}\epsilon^t_{kl}
\end{equation*}
is the probability not to detect transition from level $k$ to $l$ by a detector having total efficiency $\epsilon^t_{kl}$.
\item
The second modification is for summing-in caused by detection of all cascades between level $j$ to $i$.
The factor $a_{ji}$ in Eq.~(\ref{eqn:ctswocoi}) is modified to
\begin{equation}
a_{ji}+\sum_{k=1}^n a_{jk}a_{ki}+\sum_{k,l=1}^n a_{jk}a_{kl}a_{li}
+\cdots.
\label{eqn:sumin}
\end{equation}

\item
Finally Eq.~(\ref{eqn:ctswocoi}) should be multiplied by
\begin{equation}
\delta_{i0}
+b_{i0}
+\sum_{k=1}^n b_{ik}b_{k0}
+\sum_{k,l=1}^n b_{ik}b_{kl}b_{l0}
+\cdots
\label{eqn:sumout2}
\end{equation}
to take into account the summing-out caused by detection of a transition from the level $i$ or below.
\end{itemize}

The probability to detect the $\gamma$ transition $j \to i$ per a precursor decay is expressed by the product of Eqs.~(\ref{eqn:sumout1}), (\ref{eqn:sumin}) and (\ref{eqn:sumout2}).
If we introduce $n$-dimensional vector and matrices
$\vector{f}=\{f_p\}$,
$\matrix{x}=\{x_{ji}\}$,
$\matrix{a}=\{a_{ji}\}$,
and
$\matrix{b}=\{b_{ji}\}$,
this product is expressed by 
\begin{equation}
C_{1ji}=
\left[\vector{f}\left(\matrix{I}+\sum_{k=1}^n \matrix{b}^k\right)\right]_j 
\left[\sum_{k=1}^n \matrix{a}^k\right]_{ji}
\left[\matrix{I}+\sum_{k=1}^n \matrix{b}^k\right]_{i0},
\label{eqn:ctswcoi}
\end{equation}
where $\matrix{I}$ is the identity matrix while
$[\cdots]_j$ and $[\cdots]_{ji}$ denote
the $j$-th element of a vector and $ji$-th element of a matrix,
respectively.
Similarly Eq.~(\ref{eqn:ctswocoi}) is expressed by
\begin{equation}
C_{0ji}=
\left[\vector{f}\left(\matrix{I}+\sum_{k=1}^n \matrix{x}^k\right)\right]_j a_{ji}.
\label{eqn:ctswocoi2}
\end{equation}
Equations~(\ref{eqn:ctswcoi}) and (\ref{eqn:ctswocoi2}) can be further simplified to 
\begin{eqnarray}
C_{0ji}&=&\left[\vector{f}\matrix{X}\right]_j a_{ji},
\label{eqn:ctswocoi3}
\\
C_{1ji} &=&\left[\vector{f}\matrix{B}\right]_j 
A_{ji}
B_{i0},
\label{eqn:ctswcoi3}
\end{eqnarray}
by introducing
\begin{equation*}
\matrix{X}=\matrix{I}+\sum_{k=1}^n\matrix{x}^k,
\quad
\matrix{B}=\matrix{I}+\sum_{k=1}^n\matrix{b}^k,
\quad
\matrix{A}=\sum_{k=1}^n\matrix{a}^k.
\end{equation*}
The ratio $D_{ji}=C_{0ji}/C_{1ji}$ gives the correction factor to eliminate the coincidence-summing effect in a raw count.
\subsection{Uncertainty propagation to correction factor}
The uncertainty in the correction factor $D_{ji}$ may be propagated
from the covariances of $\{f_p\}$, $\{x_{ts}\}$, $\{\alpha_{ts}\}$,
$\{\epsilon^p_{ts}\}$ and $\{\epsilon^t_{ts}\}$.

The covariance of $f_p$ is propagated to $D_{ji}$ by
\begin{equation*}
\begin{split}
&
\quad
\sum_{p,p'=0}^n
\frac{\partial D_{ji}}{\partial f_p}
\cov(f_p,f_{p'})
\frac{\partial D_{ji}}{\partial f_{p'}}
\\
&=
\sum_{\alpha,\beta=0}^1
\sum_{p,p'=0}^n
\frac{\partial D_{ji}}{\partial C_{\alpha ji}}
\frac{\partial C_{\alpha ji}}{\partial f_p}
\cov(f_p,f_{p'})
\frac{\partial D_{ji}}{\partial C_{\beta ji}}
\frac{\partial C_{\beta ji}}{\partial f_{p'}}
\\
&=
\sum_{\alpha,\beta=0}^1
\sum_{p,p'=0}^n
\eta_{\alpha,\beta}
\left(
\frac{D_{ji}}{C_{\alpha ji}}
\frac{\partial C_{\alpha ji}}{\partial f_p}
\right)
\cov(f_p,f_{p'})
\left(
\frac{D_{ji}}{C_{\beta ji}}
\frac{\partial C_{\beta ji}}{\partial f_{p'}}
\right),
\end{split}
\end{equation*}
where 
\begin{equation}
\eta_{\alpha,\beta}=
\left\{
\begin{array}{rl}
+1 & (\alpha=\beta)\\
-1 & (\alpha\ne\beta)\\
\end{array}.
\right.
\label{eqn:eta}
\end{equation}
The covariance of $x_{ts}$ is propagated to $D_{ji}$ by
\begin{equation*}
\begin{split}
&
\quad
\sum_{t,s,t',s'=0}^n
\frac{\partial D_{ji}}{\partial x_{ts}}
\cov(x_{ts},x_{t's'})
\frac{\partial D_{ji}}{\partial x_{t's'}}
\\
&=
\sum_{\alpha,\beta=0}^1
\sum_{t,s,t',s'=0}^n
\frac{\partial D_{ji}}{\partial C_{\alpha ji}}
\frac{\partial C_{\alpha ji}}{\partial x_{ts}}
\cov(x_{ts},x_{t's'})
\frac{\partial D_{ji}}{\partial C_{\beta ji}}
\frac{\partial C_{\beta ji}}{\partial x_{t's'}}
\\
&=
\sum_{\alpha,\beta=0}^1
\sum_{t,s,t',s'=0}^n
\eta_{\alpha,\beta}
\left(
\frac{D_{ji}}{C_{\alpha ji}}
\frac{\partial C_{\alpha ji}}{\partial x_{ts}}
\right)
\cov(x_{ts},x_{t's'})
\left(
\frac{D_{ji}}{C_{\beta ji}}
\frac{\partial C_{\beta ji}}{\partial x_{t's'}}
\right),
\end{split}
\end{equation*}
and similar equations for propagation from
$\{\alpha_{ts}\}$, $\{\epsilon^p_{ts}\}$ and $\{\epsilon^t_{ts}\}$
to $D_{ji}$.

By collecting these terms,
we obtain the full equation of the uncertainty propagation to $D_{ji}$
\begin{equation}
\begin{split}
&
\quad
\displaystyle
\var(D_{ji})=
\displaystyle \sum_{\alpha,\beta=0}^1\eta_{\alpha,\beta}
\\
&
\left[
\displaystyle
\sum_{p,p'=1}^n
\left(\frac{D_{ji}}{C_{\alpha ji}}
      \frac{\partial C_{\alpha ji}}{\partial f_p}\right)
\cov(f_p,f_{p'})
\left(\frac{D_{ji}}{C_{\beta ji}}
      \frac{\partial C_{\beta ji}}{\partial f_{p'}}\right)
\right.
\\
&+
\sum_{t,s,t',s'=1}^n
\left(\frac{D_{ji}}{C_{\alpha ji}}
      \frac{\partial C_{\alpha ji}}{\partial x_{ts}}\right)
\cov(x_{ts},x_{t's'})
\left(\frac{D_{ji}}{C_{\beta ji}}
      \frac{\partial C_{\beta ji}}{\partial x_{t's'}}\right)
\\
&+
\displaystyle
\sum_{t,s,t',s'=1}^n
\left(\frac{D_{ji}}{C_{\alpha ji}}
      \frac{\partial C_{\alpha ji}}{\partial \alpha_{ts}}\right)
\cov(\alpha_{ts},\alpha_{t's'})
\left(\frac{D_{ji}}{C_{\beta ji}}
      \frac{\partial C_{\beta ji}}{\partial \alpha_{t's'}}\right)
\\
&+
\displaystyle
\sum_{t,s,t',s'=1}^n
\left(\frac{D_{ji}}{C_{\alpha ji}}
      \frac{\partial C_{\alpha ji}}{\partial \epsilon^p_{ts}}\right)
\cov(\epsilon^p_{ts},\epsilon^p_{t's'})
\left(\frac{D_{ji}}{C_{\beta ji}}
      \frac{\partial C_{\beta ji}}{\partial \epsilon^p_{t's'}}\right)
\\
&+
\left.
\displaystyle
\sum_{t,s,t',s'=1}^n
\left(\frac{D_{ji}}{C_{\alpha ji}}
      \frac{\partial C_{\alpha ji}}{\partial \epsilon^t_{ts}}\right)
\cov(\epsilon^t_{ts},\epsilon^t_{t's'})
\left(\frac{D_{ji}}{C_{\beta ji}}
      \frac{\partial C_{\beta ji}}{\partial \epsilon^t_{t's'}}\right)
\right].
\end{split}
\label{eqn:uncprop}
\end{equation}

One can easily confirm that the partial derivatives with respect to $f_p$ are
\begin{equation*}
\displaystyle
\frac{\partial C_{0ji}}{\partial f_p}=X_{pj}a_{ji},
\quad
\displaystyle
\frac{\partial C_{1ji}}{\partial f_p}=B_{pj}A_{ji}B_{i0}.
\end{equation*}

In order to calculate the partial derivative with respect to $x_{ts}$, $\alpha_{ts}$,
$\epsilon^p_{ts}$ and $\epsilon^t_{ts}$,
it is convenient to extract a part of $C_{\alpha ji}$ 
involving the transition $t \to s$ 
so that we can simplify the partial derivative such as
\begin{equation*}
\frac{\partial C_{\alpha ji}}{\partial x_{ts}}
=\frac{\partial C_{\alpha ji,ts}}{\partial x_{ts}}
\end{equation*}
.

Extraction of such terms from
Eqs.~(\ref{eqn:ctswocoi}) and (\ref{eqn:ctswcoi}) is
\begin{equation*}
\begin{split}
C_{0ji,ts}&=\left\{
\begin{array}{ll}
\displaystyle
\left[\vector{f}\matrix{X}\right]_t
x_{ts}
X_{sj}a_{ji}
&\mbox{if $t\ge j$ and $s\ge j$}\\
\\
\left[\vector{f}\matrix{X}\right]_t
a_{ts}
&\mbox{if $t=j$ and $s=i$}\\
\\
0
&\mbox{otherwise},\\
\end{array}
\right.
\\
C_{1ji,ts}&=\left\{
\begin{array}{ll}
\displaystyle
\left[\vector{f}\matrix{B}\right]_t 
b_{ts}
B_{sj}
A_{ji}
B_{i0}
&\mbox{if $t\ge j$ and $s\ge j$}\\
\\
\displaystyle
\left[\vector{f}\matrix{B}\right]_j 
A'_{jt}
a_{ts}
A'_{si}
B_{i0}
&\mbox{if $j\ge t\ge i$ and $j\ge s \ge i$}\\
\\
\displaystyle
\left[\vector{f}\matrix{B}\right]_j 
A_{ji}
B_{it}
b_{ts}
B_{s0}
&\mbox{if $t \le i$ and $s \le i$}\\
\\
0 \quad
&\mbox{otherwise},\\
\end{array}
\right.
\label{eqn:ctswcoidel}
\end{split}
\end{equation*}
where
\begin{equation*}
\matrix{A}^\prime=\matrix{I}+\sum_{k=1}^n\matrix{a}^k.
\end{equation*}
The partial derivatives of $C_{\alpha ji}$ for a parameter 
($x_{ts}, \alpha_{ts}, \epsilon^p_{ts},$ and $\epsilon^t_{ts}$)
can be easily calculated by using these equations.

\section{Example: $^{134}$Cs decay $\gamma$ detection}
We applied our formulation of the deterministic uncertainty propagation to the total and peak efficiencies of $^{134}$Cs decay $\gamma$-rays detection by a coaxial HPGe detector of EC-JRC IRMM.
The full-energy peak and total efficiencies in the energy 60-1400~keV on the top of the detector were determined by interpolation of reference data with polynomial logarithmic function~\cite{VS2006}.
The experimental efficiencies were determined using a set of $^{241}$Am, $^{109}$Cd, $^{139}$Ce, $^{51}$Cr, $^{137}$Cs, $^{54}$Mn and $^{65}$Zn monoenergetic standard sources supplied by LEA Laboratoire Etalons d’Activit\'{e}, France.
The efficiencies at 122 keV and 136.5 keV were determined using a $^{57}$Co standard source considering coincidence-summing effects as negligible for the p-type detector used in this work.
An accurate detector model was developed by optimization of Monte Carlo simulations using {\sc MCNP} code~\cite{JFB1997} based on measured efficiencies at various positions around the detector.
To extend the energy range for the polynomial fitting above the range covered by the ``single-line'' standard sources the efficiencies at 1173.2, 1274.53, and 1332.5~keV determined by MCNP calculations were used.
The detector response for these energies was optimized by measurements at 10.5~cm source-to-detector distance and calculated for close geometry to avoid influence of the coincidence-summing effects.
The covariance matrices of the $^{134}$Cs decay $\gamma$-ray efficiencies were determined following the method described in Ref.~\cite{LPG1990}.
%

We calculated the coincidence-summing correction factor,
and also propagated the covariances of $\epsilon^t_{ji}$ and $\epsilon^p_{ji}$ as well as uncertainties in $f_p$, $x_{ji}$ and $\alpha_{ji}$~\cite{MMB2013} to the uncertainty in the correction factor.
\begin{figure}[h]
\centering
\includegraphics[width=1.0 \columnwidth]{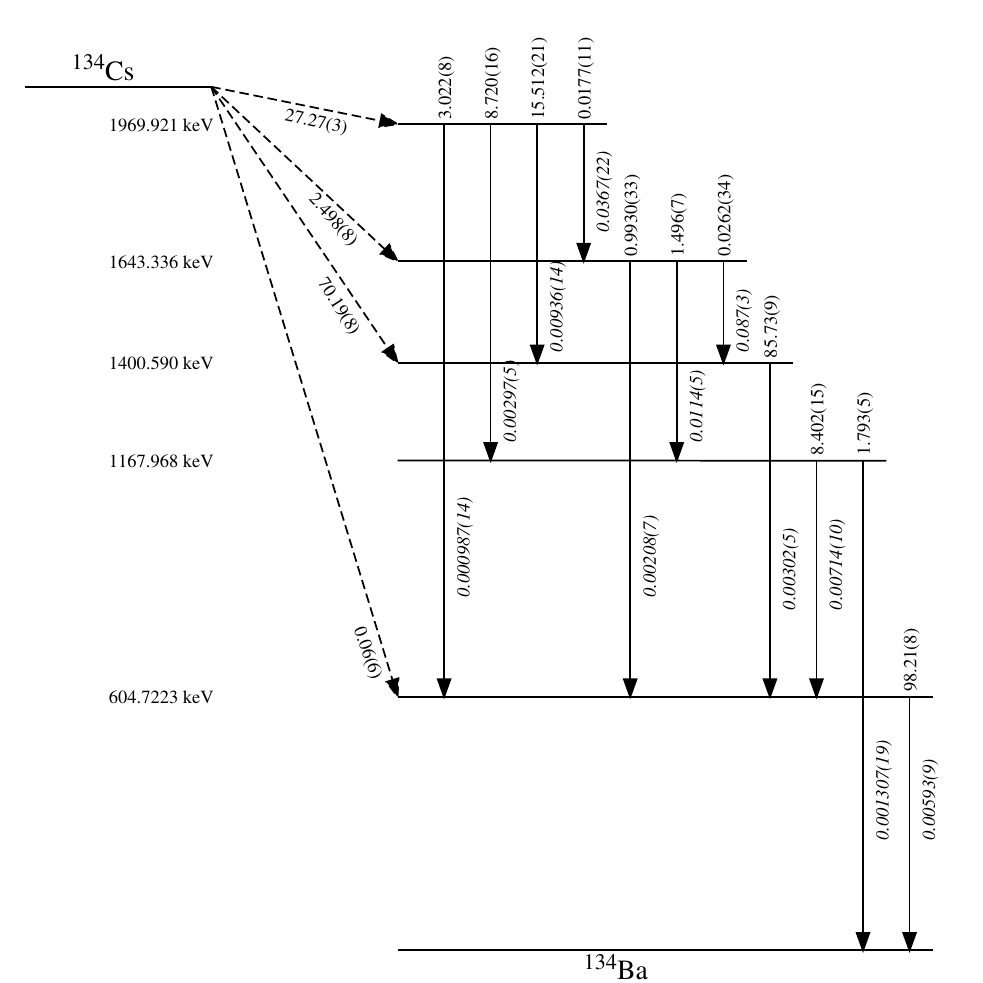}
\caption{$\beta$ transition and $\gamma$ transition+internal conversion probabilities in $^{134}$Cs $\beta^-$ decay~\cite{MMB2013}. Note that these $\gamma$ transition probabilities are before normalization according to Eq.~(\ref{eqn:norm}).
}
\label{fig:cs134}
\end{figure}
The off-diagonal elements of the decay data covariances are not available in Ref.~\cite{MMB2013}, and we simply ignored the correlations in the decay data.
\begin{table*}
\centering
\caption{Total and partial uncertainties in the coincidence-summing correction factor $D_{ji}=C_{0ji}/C_{1ji}$.
The negative correlation between $C_{0ji}$ and $C_{1ji}$ is taken into account (\ie $\eta_{\alpha,\beta}=-1$ when $\alpha\ne\beta$) in the case of ``full correlation case'',
while its is ignored (\ie $\eta_{\alpha,\beta}=0$ when $\alpha\ne\beta$) in the case of ``without negative correlation''.
See  Eq.~(\ref{eqn:eta}) for the definition of $\eta_{\alpha,\beta}$.
}
\label{tab:result}
\begin{tabular}{cccccccccccccccc}
\hline
\multirow{2}{*}{$E_\gamma$}&
\multirow{2}{*}{$j\to i$}&
\multirow{2}{*}{$D_{ji}$}&
\multicolumn{6}{c}{$\Delta D_{ji}$ with full correlation (\%)}&
&
\multicolumn{6}{c}{$\Delta D_{ji}$ without negative correlation (\%)}\\
\cline{4-9}
\cline{11-16}
& & &
 Total& $f$ & $x$ & $\alpha$ & $\epsilon^p$ & $\epsilon^t$&
&
 Total& $f$ & $x$ & $\alpha$ & $\epsilon^p$ & $\epsilon^t$\\
\hline
  242.7 & 4 $\to$ 3 & 1.173 & 0.92 & 0.00 & 0.13 & 0.00 & 0.00 & 0.91 &  & 18.47 & 0.45 & 18.35 & 0.39 & 1.80 & 0.91 \\
  326.6 & 5 $\to$ 4 & 1.225 & 1.04 & 0.00 & 0.35 & 0.00 & 0.00 & 0.98 &  &  9.00 & 0.16 &  8.80 & 0.30 & 1.57 & 0.98 \\
  475.4 & 4 $\to$ 2 & 1.162 & 0.92 & 0.00 & 0.17 & 0.00 & 0.00 & 0.90 &  &  1.72 & 0.45 &  0.69 & 0.07 & 1.22 & 0.90 \\
  563.2 & 2 $\to$ 1 & 1.175 & 0.88 & 0.00 & 0.08 & 0.00 & 0.00 & 0.88 &  &  1.35 & 0.15 &  0.36 & 0.01 & 0.95 & 0.88 \\
  569.3 & 5 $\to$ 3 & 1.173 & 0.92 & 0.00 & 0.13 & 0.00 & 0.00 & 0.91 &  &  1.24 & 0.16 &  0.23 & 0.02 & 0.79 & 0.91 \\
  604.7 & 1 $\to$ 0 & 1.100 & 0.56 & 0.01 & 0.00 & 0.00 & 0.00 & 0.56 &  &  0.94 & 0.15 &  0.18 & 0.01 & 0.72 & 0.56 \\
  795.9 & 3 $\to$ 1 & 1.104 & 0.70 & 0.00 & 0.08 & 0.00 & 0.00 & 0.70 &  &  1.03 & 0.14 &  0.17 & 0.01 & 0.72 & 0.70 \\
  802.0 & 5 $\to$ 2 & 1.161 & 0.91 & 0.00 & 0.17 & 0.00 & 0.00 & 0.90 &  &  1.27 & 0.16 &  0.31 & 0.01 & 0.83 & 0.90 \\
 1038.6 & 4 $\to$ 1 & 1.033 & 0.65 & 0.00 & 0.09 & 0.00 & 0.07 & 0.64 &  &  1.29 & 0.45 &  0.47 & 0.01 & 0.90 & 0.64 \\
 1168.0 & 2 $\to$ 0 & 0.921 & 0.54 & 0.00 & 0.05 & 0.00 & 0.19 & 0.50 &  &  1.14 & 0.15 &  0.44 & 0.00 & 0.92 & 0.50 \\
 1365.2 & 5 $\to$ 1 & 0.874 & 0.71 & 0.00 & 0.10 & 0.00 & 0.28 & 0.64 &  &  1.83 & 0.16 &  0.35 & 0.00 & 1.67 & 0.64 \\
\hline
\end{tabular}
\end{table*}

Table~\ref{tab:result} summarizes the correction factors with their total and partial uncertainties.
The partial uncertainties originated from the uncertainty in $f$, $x$, $\alpha$, $\epsilon^p$ and $\epsilon^t$ correspond to the square root of the five terms on the right-hand side of Eq.~(\ref{eqn:uncprop}).

If we do not take into account the correlation between $C_{0ji}$ and $C_{1ji}$ in the uncertainty propagation to $D_{ji}$ (\ie $\eta_{\alpha\beta}=0$ instead of -1 when $\alpha\ne\beta$),
large total uncertainties are seen for two transitions, 4$\to$3 and 5$\to$4,
which reflects the large uncertainties in the $\gamma$+internal conversion probabilities of these transitions ($\Delta x_{ji}/x_{ji}$=13\% for 4$\to$3 and 6\% for 5$\to$4).
However, the contribution of the uncertainties in the nuclear data and peak efficiencies are drastically reduced if we take into account the negative correlation (\ie $\eta_{\alpha\beta}=-1$ when $\alpha\ne\beta$).
This cancellation is not seen for the uncertainty in the total efficiency $\epsilon^t$ since the true count $C_{0ji}$ does not depend on the total efficiency.

\section{Summary}
We have developed an analytic approach to propagate the covariances of the nuclear data and detector efficiencies to the uncertainty in the coincidence-summing correction factor,
and applied it to detection of $^{134}$Cs decay $\gamma$-lines by a HPGe detector.
We demonstrated that the uncertainty in the correction factor is significantly reduced when we take into account correlation better the counts with and without coincidence summing appropriately.

%
%

\end{document}